\documentclass[conference]{IEEEtran}
\usepackage{cite}
\usepackage{amsmath,amssymb,amsfonts}
\usepackage{algorithmic}
\usepackage{graphicx}
\usepackage{textcomp}
\usepackage{xcolor}
\usepackage[obeyspaces]{url}
\def\BibTeX{{\rm B\kern-.05em{\sc i\kern-.025em b}\kern-.08em
    T\kern-.1667em\lower.7ex\hbox{E}\kern-.125emX}}

\usepackage{flushend}

\newif\iffinal

\finaltrue

\iffinal
  \newcommand{\yadu}[1]{}
  \newcommand{\TODO}[1]{}
  \newcommand{\asv}[1]{}
  \newcommand{\kyle}[1]{}
  \newcommand{\katznote}[1]{}

\else
  \newcommand{\TODO}[1]{{\textcolor{red}{ TODO: #1 }}}
  \newcommand{\yadu}[1]{{\textcolor{orange}{ Yadu: #1 }}}
  \newcommand{\asv}[1]{{\textcolor{purple}{ ASV: #1 }}}
  \newcommand{\kyle}[1]{{\textcolor{blue}{ Kyle: #1 }}}
  \definecolor{darkgreen}{rgb}{0,0.6,0}
  \newcommand{\katznote}[1]{{\textcolor{darkgreen}{ Dan: #1 }}}
\fi

\newif\ifchange

 \newcommand{\name}[1]{{ImSim Workflow}}

\begin{document}

\title{Extreme Scale Survey Simulation with Python Workflows\\
\thanks{For the LSST Dark Energy Science Collaboration}
}

\author{\IEEEauthorblockN{Antonia Villarreal\IEEEauthorrefmark{1}, Yadu Babuji\IEEEauthorrefmark{2}, Tom Uram\IEEEauthorrefmark{1}, Daniel S. Katz\IEEEauthorrefmark{3}, Kyle Chard\IEEEauthorrefmark{2}, and Katrin Heitmann\IEEEauthorrefmark{4}\\
{(The LSST Dark Energy Science Collaboration)}}

\IEEEauthorblockA{\IEEEauthorrefmark{1}Argonne Leadership Computing Facility, Argonne National Laboratory, Argonne, Illinois 60439} %\\Emails: avillarreal@anl.gov, turam@anl.gov}
\IEEEauthorblockA{\IEEEauthorrefmark{2}Department of Computer Science, University of Chicago, Chicago, Illinois 60637} %\\Emails:stuffhere\\ORCID: stuffhere}
 \IEEEauthorblockA{\IEEEauthorrefmark{3}NCSA \& CS \& ECE \& iSchool, University of Illinois, Urbana, Illinois} 
 %ORCID:0000-0001-5934-7525}
\IEEEauthorblockA{\IEEEauthorrefmark{4}High Energy Physics Division, Argonne National Laboratory, Argonne, Illinois 60439}} % Email: heitmann@anl.gov}}

\maketitle

\begin{abstract}
The Vera C. Rubin Observatory Legacy Survey of Space and Time (LSST) will soon carry out an unprecedented wide, fast, and deep survey of the sky in multiple optical bands. The data from LSST will open up a new discovery space in astronomy and cosmology, simultaneously providing clues toward addressing burning issues of the day, such as the origin of dark energy and and the nature of dark matter, while at the same time yielding data that will, in turn, pose fresh new questions. To prepare for the imminent arrival of this remarkable data set, it is crucial that the associated scientific communities
be able to develop the software needed to analyze it. Computational power now available allows us to generate synthetic data sets that can be used as a realistic training ground for such an effort. This effort raises its own challenges---the need to generate very large simulations of the night sky, scaling up simulation campaigns to large numbers of compute nodes across multiple computing centers with different architectures, and optimizing the complex workload around memory requirements and widely varying wall clock times. 
We describe here a large-scale workflow that melds together Python code to steer the workflow, Parsl to manage the large-scale distributed execution of workflow components, and containers to carry out the image simulation campaign across multiple sites. Taking advantage of these tools, we developed an extreme-scale computational framework and used it to simulate five years of observations for 300 square degrees of sky area. We describe our experiences and lessons learned in developing this workflow capability, and highlight how the scalability and portability of our approach enabled us to efficiently execute it on up to 4000 compute nodes on two supercomputers. 

\end{abstract}


\section{Introduction}
\label{section:introduction}

Developing any scientific instrument combining hardware and software can be challenging, as there is often the need to design, build, and test the software while the hardware is incomplete. One method to partially address this challenge is to create synthetic data that mimics what will be captured by the instrument and can then be used to develop and test analysis
software. 
This is the situation facing the Vera C. Rubin Observatory Legacy Survey of Space and Time (LSST), an astronomical survey that will enable cutting-edge science through analysis of its data releases. The LSST Dark Energy Science Collaboration (DESC) is preparing for these data releases to transform them into groundbreaking scientific results, with a particular focus on understanding of the evolution of the universe and its accelerating expansion rate (the ``dark energy'' problem). 

In order to make best use of the data, extensive development, testing, and validation of the scientific analysis software is required. To this end, DESC has developed data challenges: data releases of simulated data in increasing steps of complexity for the purposes of developing scientific pipelines ahead of LSST data releases~\cite{2020MNRAS.497..210S,Abolfathi_2021}. 
We describe here the computational process, using Python workflows across two
high performance computing (HPC) facilities, to create the image simulations for the recent Data Challenge~2 (DC2). 

While the task of generating simulated data is not uncommon in scientific experiments, DC2 posed unique computational challenges. The most significant hurdles were the need to develop an easy-to-understand description of the workflow, to orchestrate a simulation campaign using research software with a range of dependencies, to manage the extreme-scale
computational requirements, to port the workflow between two supercomputers, and to optimize performance such that allocations at the facilities were efficiently used. 

The DESC aims to understand the evolution of the universe by exploring the LSST data via multiple cosmological probes, as discussed in detail in the LSST DESC Science Requirements Document (SRD)~\cite{thelsstdarkenergysciencecollaboration2018lsst}. These probes will allow us to measure the behavior of phenomena such as cosmic acceleration to new levels of precision---provided that any potential sources of systematic error are well characterized, understood, and mitigated. In order to test the robustness of the scientific pipelines, DESC can benefit from a large simulated data volume that represents a partial realization of the full ten-year survey. For details on the entirety of this effort, the LSST DESC present a description of the complex process of creating this data volume from numerical simulations of the large-scale structure in the universe to processed simulated images~\cite{Abolfathi_2021}. We focus our discussion here on a specific workflow portion within that greater effort---that of the Python-based \name{} used to simulate images comparable to that of the real Rubin Observatory LSST Camera.

The \name{} uses the image simulation code, {\tt imSim}, to simulate each of the camera's 189 CCD sensors. 
{\tt imSim} takes as input an ``instance catalog,'' which consists of millions of individual entries of commands for the code to simulate specific objects as seen by the camera. 
For each sensor, {\tt imSim} must determine which objects cast light on that specific sensor. For each command, the underlying libraries must go through a complicated process of determining which computational method is most suitable for the specific object to be drawn, which varies both the memory utilized and the compute time on a per object basis. Each simulated sensor represents a task that is independent of any other simulated sensor, though it may share common information within a given telescope observation.
 
We develop the \name{} using the Parsl parallel scripting library to encode
the workflow in Python and manage the large-scale simulation campaign. 
Briefly, the workflow is given a set of instance catalogs and a target 
area for simulation over time. The workflow must then
identify the observations to be simulated, estimate the amount of work that can fit on a single compute node, determine all commands that need to be run with {\tt imSim}, and distribute these tasks across many compute nodes. 
To address the challenges deploying the evolving imSim code on two supercomputers
we adopt containers as a way of encapsulating a point-in-time software environment
for the ecosystem and for porting it between systems.
We have used the \name{} to simulate five years of observations for a 
300 square degree area of the night sky. This campaign consumed approximately
100 million core hours across the NERSC Cori and ALCF Theta supercomputers.

We describe the ambitious simulation program in the rest of this paper as follows. In Section~\ref{section:campaign}, we outline the scientific motivation of the image simulation campaign, its computational requirements, and the tools that we used. In Section~\ref{section:workflow}, we describe the \name{}. 
In Section~\ref{section:experiences} we provide a brief discussion about the choices we made when developing the workflow before we present our results in Section~\ref{section:results}.
In Section~\ref{section:discussion}, we show various diagnostics regarding our computational approach and discuss the problems encountered during the computational campaign in order to determine what improvements may be made on future Data Challenges. In Section~\ref{section:related} we comment on related work and, finally, we summarize our work in Section~\ref{section:conclusions}. 

\section{Image Simulation Campaign}
\label{section:campaign}

We first describe the goals of our work and present relevant background information about the underlying  {\tt imSim} package.

\subsection{Data Challenge 2}
\label{section:DC2}
The goal of DC2 was to generate simulated images that can be processed with Rubin's LSST Science Pipelines as though they were true survey images. The \name{} that we describe here is one part of a long chain of steps, which is described in further detail in the DC2 paper~\cite{Abolfathi_2021}. In brief, large-scale cosmological simulations are run to create one realization of a possible universe. Next, we identify where galaxies would exist inside the simulation and create catalogs with their properties~\cite{cosmodc2}. We then generate ``instance catalogs'' --- for a given observation of the night sky, we create a list of objects that would be observed by the LSST Camera (these include Milky Way stars as well as extragalactic sources). The instance catalogs are then used as input to {\tt imSim}, which in turn will generate an image for each object in the instance catalog. The resulting images can then can be processed with the LSST Science Pipelines.

\begin{figure}[h]
  \includegraphics[width=0.97\columnwidth]{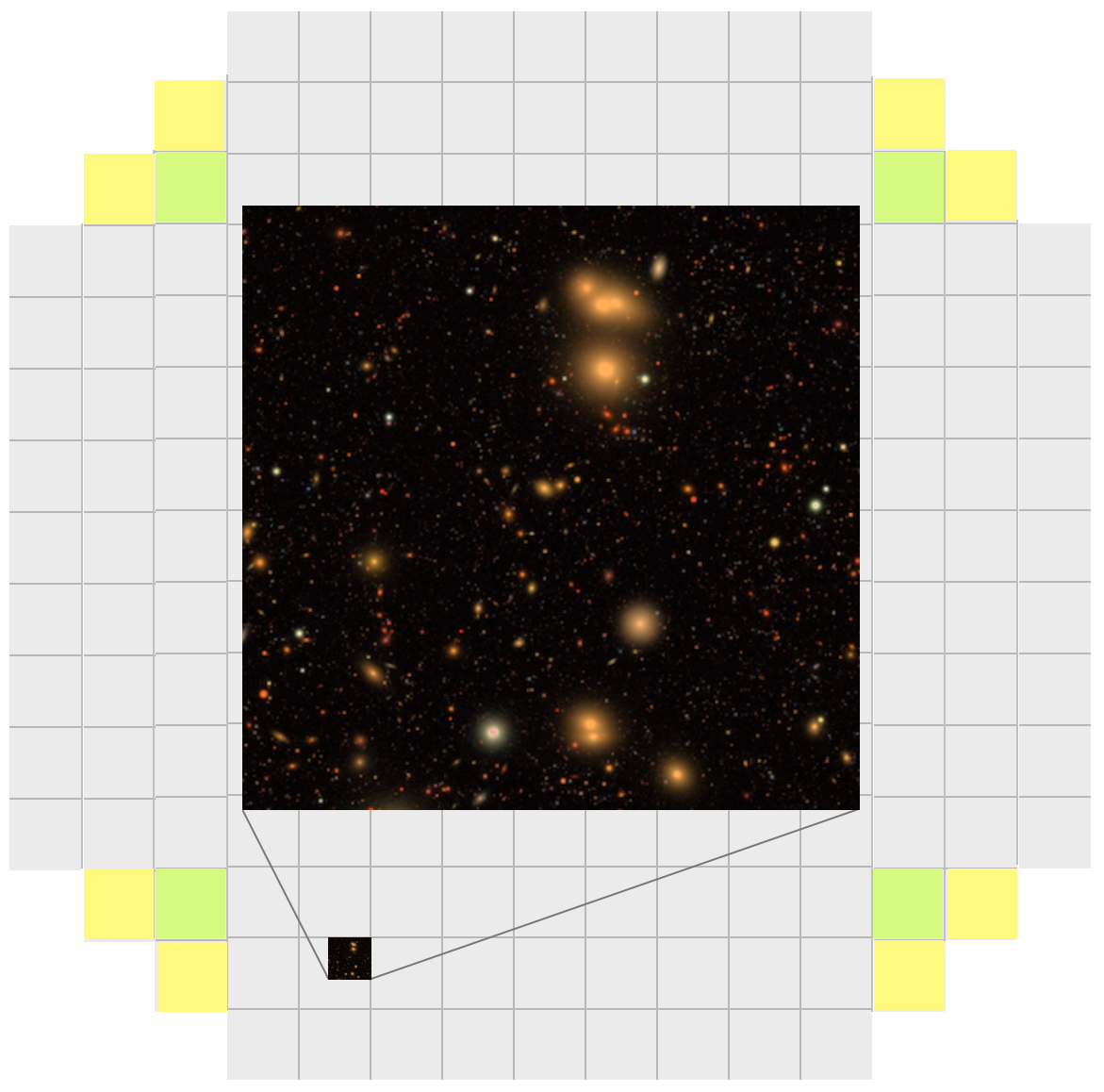}
    \vspace{-0.1in}
  \caption{Final processed images are overlaid over the LSST camera focal plane of 189 individual sensors (light gray squares). The yellow and green squares are wavefront and guide sensors.}
  \label{fig:focalplane}
    \vspace{-0.1in}
\end{figure}

The survey volume simulated for DC2 is a five-year span over 300 square degrees of sky, for a ``wide-fast-deep'' (WFD) scan, with a 1 square degree insert region that serves as the ``deep drilling field'' (DDF). Each observation, referred to as a \textbf{visit}, represents a single telescope pointing (with a total duration of 30 seconds). Each visit further consists of 189 \textbf{sensors}, distributed across the camera as seen in Figure~\ref{fig:focalplane}. After simulating five years of observations, these images can then be stacked on top of each other and processed to create images such as those seen in the figure. Specific details regarding the distribution of visits in the survey volume are provided in the DC2 paper~\cite{Abolfathi_2021}.

Note that the above volume sets an upper limit on the scale of the data---with over 30,000 visits each containing up to 189 sensors, we would need to simulate on the order of five million images. In practice, some of these images have no objects in them and may be skipped, but the total volume of image data generated is approximately 100 TB. The sheer number of independent images and the computational requirements, particularly with respect to memory usage, represent the major difficulty in this simulation effort, as detailed below.

\subsection{imSim}
\label{section:imsim}
The image simulations are carried out using the {\tt imSim} software package. This package uses the  {\tt Galsim} software library to carry out rendering of stars and extragalactic objects as they would be observed with the 3.25 Gigapixel LSST Camera \cite{rowe2015galsim}. The image simulations include the impact of atmosphere, optics, and sensor effects on the observed images. The underlying software utilizes C++ for the bulk of its calculations, while Python provides a flexible user interface that can be modified to many specific tasks. In addition, {\tt imSim} takes advantage of some functionality provided via the LSST Science Pipelines.

To better understand the exact requirements on the computational workflow, we briefly discuss the nature of the inputs and outputs connected to {\tt imSim}. As stated above, from numerical simulations we arrive at the information regarding a single pointing of a telescope for a single visit---what we refer to as an \textbf{instance} catalog. This instance catalog contains a list of entries---one for each star, galaxy, or dynamic object. Each entry contains information regarding the location of the object on the focal plane, the shape of the object in advance of atmospheric and sensor effects, the position of the telescope to enable the inclusion of the effects from the relative airmass, and more. For each of the 189 sensors, we can then request an output image that will correspond to the output generated by the real telescope---albeit for our artificial universe.

The {\tt imSim} software can be passed a list of sensors to be simulated concurrently. It can distribute this work across the cores on a single node using the Python {\tt multiprocessing} library. The master {\tt imSim} process then shares information regarding the atmospheric screens to each of these tasks, minimizing a substantial memory overhead. As each individual visit requires distinct atmospheric screens, this becomes a computational limitation, as packing many distinct visits onto a single compute node involves calculating many atmospheric screens---each screen taking more memory than several additional sensors. The remaining computational barrier is then the number of sensors that one wishes to simulate simultaneously on a given compute node. While atmospheric screens are shared, drawing objects simultaneously can result in overlap of objects with large memory use. Planning around this is exceedingly complicated---{\tt Galsim} has a complex decision tree for determining the best algorithm for drawing an object on the sensor and this is not inherently easy to determine at the instance catalog level. As such, there is a potential for many large memory footprint objects to overlap during simulation that can lead to excessive memory use on a node. The version of {\tt imSim} used by the \name{} during our DC2 campaign did not convey any information regarding expected memory footprint back to the master process constituting another design constraint to our workflow.

\section{The \name{}}
\label{section:workflow}

As described above, the simulation size (upwards of five million independent images to be generated) and the associated resource needs (on a per image level) are the ultimate challenges that need to be addressed by the use of a sophisticated, large-scale workflow. 
The computational barriers to overcome are as follows:
\begin{itemize}
    \item An individual visit contains up to 189 independent computational processes.
    \item The memory footprint of each compute thread requires splitting visits across multiple compute nodes.
    \item The memory requirements of each thread vary considerably depending on the simulated objects.
    \item Each individual visit may have varying compute times, with a scatter that can be different from sensor to sensor on that visit.
    \item The compute time of all visits greatly exceeds the available wall-clock times for a single job on most submission systems.
    \item Given the above, and the fact that some tasks may fail, each individual task needs to be tracked for completion and restarts from checkpoint files need to be enabled.
    \item Additionally, given the large compute time needed, it is important that the workflow can be moved dynamically between computing centers based on available capacity.
\end{itemize}

To meet these requirements, we developed the \name{} using a combination of the Parsl parallel scripting library and containers.  
We rely on Parsl as a way of encoding the workflow in Python 
and for efficiently managing the execution of our workload on HPC
resources. We use containers to capture the complex software
environment and provide portability across systems and architectures.
We begin by describing Parsl and containers before outlining the \name{}.

\subsection{Parsl}
\label{section:parsl}
Parsl~\cite{babuji2019} is a flexible and scalable parallel programming library for Python. Parsl offers simple constructs for encoding parallelism in standard Python programs. Specifically, developers can annotate Python functions to specify opportunities for concurrent execution. These annotated functions, called apps, may represent pure Python functions or calls to external applications. Parsl further allows invocations of these apps, called tasks, to be connected together into a dependency graph by shared input/output data flow (e.g., Python objects or files). 
Parsl uses the dependency graph to safely manage concurrent execution of tasks. 

Parsl leverages an extensible and scalable runtime system that allows it to efficiently execute tasks on many cores and processors.  Parsl supports various target resources including clouds (e.g., Amazon Web Services and Google Cloud), clusters (e.g., using Slurm, Torque/PBS, HTCondor, Cobalt), and container orchestration systems (e.g., Kubernetes). It also supports various executors that manage how tasks are executed on computing resources (e.g., using a distributed MPI fabric or managing task execution via a pilot job model). In most cases, Parsl executors deploy a \textit{worker} agent to each provisioned node and these agents are then responsible for executing tasks passed from the Parsl runtime via the executor. Parsl programs can scale from several cores on a single computer through to hundreds of thousands of cores across many thousands of nodes on a supercomputer.

Parsl provides the backbone for the \name{} as it provides several
crucial capabilities. Most notably, Parsl allows us to write the 
entire \name{} in Python---a language well-understood by the 
DESC community. It also allows us to scale the workload to address
the performance needs outlined above, can efficiently
utilize thousands of nodes on leadership-class supercomputers, 
supports the deployment and invocation of code in containers, 
provides checkpointing and fault tolerance, and offers logging
and monitoring capabilities to diagnose errors.

\subsection{Containers}
\label{section:docker}

As discussed in Section~\ref{section:imsim}, there is a significant amount of underlying software required to run {\tt imSim} and this is further complicated by the fact that {\tt imSim} was under active development while we developed the \name{}.
In practice, keeping the {\tt imSim} software updated and functional on all compute nodes is a nontrivial
exercise, especially ensuring that it functions identically across multiple compute resources. 
In order to accomplish this task, we rely upon the use of containerization software. 

Our target computing resources at ALCF and NERSC support different container technology (Singularity 
and Shifter, respectively). To address the need to deploy software on many computing resources
we first created a Docker container as a base representation of the environment 
and we used it to create Shifter and Singularity images as needed. 

The Docker image we use consists of a base layer that contains the LSST Science Pipelines, followed by additional commands which copy specifically tagged versions of all underlying software needed for image simulation.

\subsection{\name{} Implementation}
\begin{figure}[h]
  \includegraphics[width=0.97\columnwidth]{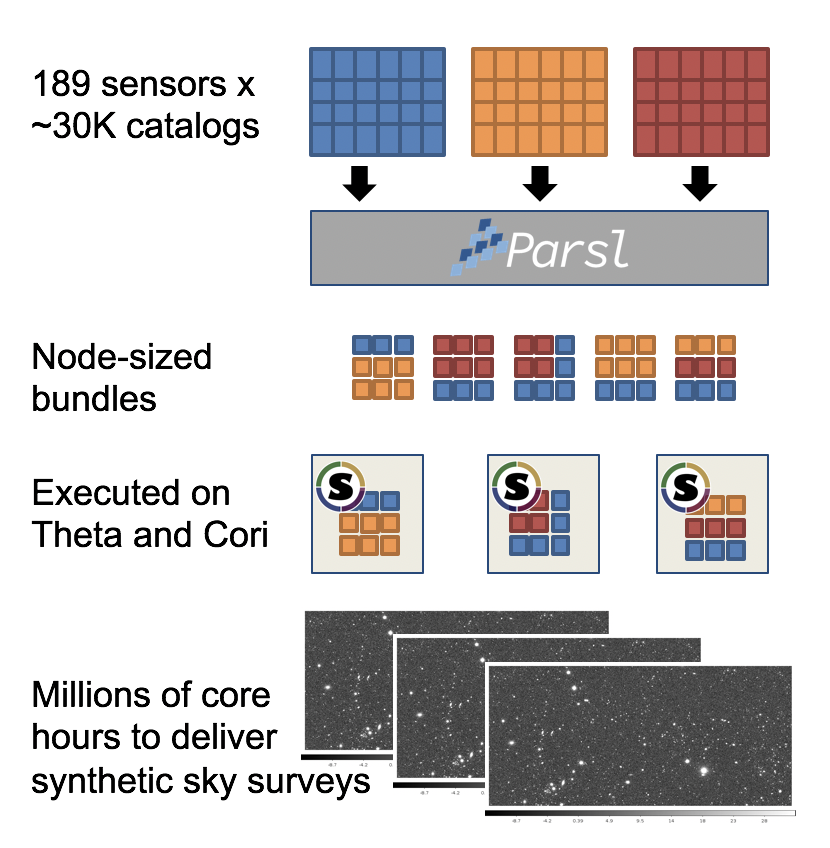}
    \vspace{-0.1in}
  \caption{A visual diagram of the workflow from the instance catalog level to outpupt images.}
  \label{fig:imsimlayout}
    \vspace{-0.1in}
\end{figure}

We implemented the \name{} using a single Python driver with Parsl apps
for each of the {\tt imSim} commands. The workflow is invoked with pointers to
input instance catalogs, a target area of the sky, and a range of
configuration parameters to run the simulation.  A high level view 
of the workflow is shown in Figure~\ref{fig:imsimlayout}. 
The workflow is available on GitHub: 
\url{https://github.com/LSSTDESC/DESC_DC2_imSim_Workflow}.

The workflow combines various stages, each implemented as its
own Parsl app, such as
to transfer instance catalog data, process instance catalogs
and define simulation tasks, bundle collections of tasks for efficient resource
usage, run the imSim code on the allocated bundle of tasks, determine which
tasks produced output files, and collect output files for storage.  
The control logic for the workflow is implemented in approximately 
100 lines of Python code (\texttt{parsl-driver.py}). Each of the Parsl apps
are implemented in separate Python scripts that perform the defined
operation. 

Each of the Parsl apps is executed within a container. The 
workflow allows us to seamlessly switch between container technology using
a simple configuration flag. This flag informs Parsl which container set-up is supported
by the computing facility, pulls containers from online storage in the format necessary 
for the software, and then sets a predefined wrapper. This wrapper consists of either a Shifter or Singularlity command that will load a container onto a compute node and execute a set of Python tasks in order to start simulating the visits that have been packaged in that specific work bundle. Since the underlying Python command does not interact with the container directly, changing this chosen wrapper is all that is necessary to port between different computing centers.

Each of the Parsl apps is run with checkpointing enabled. This allows
the \name{} to recover from failures and in particular to save the state
of app execution in the cases where nodes fail or allocation wall time
is exceeded. 

\section{Experiences developing the \name{}}
\label{section:experiences}
We briefly describe choices made when developing the workflow and deploying it on large-scale compute
resources. 

\subsection{Deploying the workflow}
The Parsl model allows us to acquire compute resources in different ways.   
We explored two main methods: deploying Parsl on the login node and submitting Parsl as a job for 
execution on the launch node. 

Each option brings its own advantages and disadvantages.

We first explored the most common deployment model to initialize the main Parsl driver on the login or workflow nodes on the HPC resource, which we refer to as ``remotely driven''. In this case, the driver submits the actual scheduler job request, asking for a certain amount of compute resources to accommodate Parsl workers, and then (when the job launches) transfers tasks from the driver onto the compute resources for calculations to be run.  Here the main advantage is that the driver is able to coordinate multiple submission blocks simultaneously; this can be a significant advantage on some scheduling systems which favor smaller jobs to fill utilization gaps. Potentially, if the remote driver has sufficient access, it may even be able to leverage resources from multiple HPC sites simultaneously. However, this relies strongly on the stability of the resource on which the driver runs---if the driver is brought offline for any reason during the time at which a requested job launches, the workers will be unable to get the information about their tasks, wasting valuable computation time.

We also explored submitting the Parsl driver for execution on the compute or launch node for the HPC system, which we  refer to as ``locally driven''. Configured in this way, scheduler submission scripts include the steps to initialize Parsl on a single compute or launch node, which then initializes a large number of workers across the remaining compute nodes and begins to assign tasks accordingly to these workers. This has the advantage of being entirely local to the compute resources, which reduces communication needs; further, it removes the need to have a potentially long-running driver consume resources on an external node. The main disadvantage of this approach is that it is more difficult to run multiple job submission blocks simultaneously, as there will be no block to block communication.

\subsection{Extreme-scale task execution}
\begin{figure}[h]
  \includegraphics[width=0.97\columnwidth]{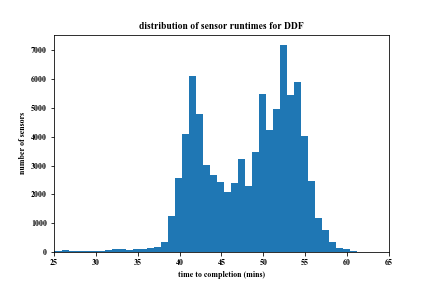}
    \vspace{-0.1in}
  \caption{The distribution of image simulation times for year three of the DDF. The variation on the order of 10\% of the total walltime is present across all simulated images.}
  \label{fig:visitdist}
    \vspace{-0.1in}
\end{figure}

Once resources have been acquired, Parsl workers can be initialized across resources via the use of one of Parsl's
executors. 
\name{} exhibits the unique requirement of needing to run both local calculations (e.g., to bundle tasks) as well
as remote computations (e.g., to invoke {\tt imSim}).  To address this requirement we use Parsl's
{\tt ThreadPoolExecutor} and {\tt HighThroughputExecutor} concurrently. 
We use the former to manage local calculations regarding the bundling of individual image simulation commands into groupings for a single Parsl worker to operate on.  Here Parsl creates a pool of threads 
that are run alongside the driver program and can share memory between one another and the 
driver script. 
The latter is an executor packaged with Parsl designed for managing the execution of Parsl apps on thousands of nodes---allowing us to fully exploit supercomputers such as Theta.  The {\tt HighThroughputExecutor} configuration begins a multiprocessing driven pool of workers across compute resources, runs an interchange that allows Parsl to divide the work among workers, and manages tracking the completion of Parsl apps and the introduction of additional Parsl apps as resources become available again. This last point in particular allows us to work around a key computational barrier listed above: the fact that individual visits have widely varying compute times, with some visits taking hours longer than others. Figure~\ref{fig:visitdist} demonstrates how even for the shorter computational task of the DDF, image output times can vary considerably compared to total wallclock time. With this configuration, short tasks are completed and new work can be placed into the newly freed workers, allowing us to greatly increase the efficiency of our computational resource use. We discuss how this plays out below.

\subsection{Task Bundling}
\label{section:bundling}

After testing with early versions of \name{}, we observed periods of inefficient
resource utilization. This inefficiency was primarily caused by the varied resource usage of 
individual tasks.  To improve efficiency we developed a bundling process that would identify
chunks of work that could be run concurrently on a single node. 
The first step in the bundling process is a pre-processing step to identify all sensors lying in the target region on a visit-by-visit basis. This is necessary to avoid tasks that take place outside our underlying simulated volume; the large focal plane of the camera can sometimes extend over this artificial boundary. From the list of visits and sensors that will need to be simulated, we then take rough estimates of the memory footprint of the individual components---the memory of the Docker container, the memory of shared resources for a visit, and the expected peak memory use of an individual sensor---and compare that against the available resources of the computing resource. The bundling algorithm then attempts to group simulation tasks together in a logical way, following several rules:
\begin{itemize}
    \item For each compute node, prioritize sensors from a single visit to optimize the use of shared resources.
    \item For each compute node, limit thread use such that peak memory use will not result in exceeding the available memory on the node.
    \item Further limit thread use to the number of threads on a given compute node---initial testing suggested hyperthreading would be ineffective for our use cases.
\end{itemize}
\begin{figure}[h]
  \includegraphics[width=0.97\columnwidth]{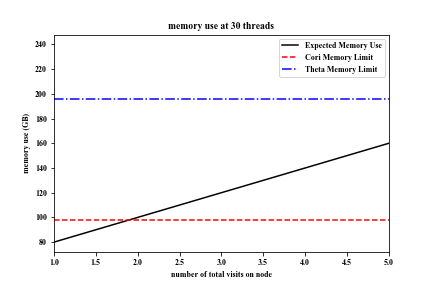}
    \vspace{-0.1in}
  \caption{The memory footprint on a given compute node for a set number of sensors but an increasing number of independent visit tasks. The solid black line shows the expected memory use, while the dashed (dot-dashed) line shows the hardware limitations at Cori (Theta).}
  \label{fig:memoryusepervisit}
    \vspace{-0.1in}
\end{figure}
Each of these particular rules represents a necessary compromise between optimal use of the compute resources available and the physical limitations of those same resources. One example is shown in Figure~\ref{fig:memoryusepervisit}, which demonstrates how the total number of visits on a node changes the memory usage---even with the number of compute threads kept stable. This is the result of inefficient use of shared resources, specifically atmospheric screens that are identical for a given observation---providing a strong incentive to minimize the number of independent visits on a given compute node. As a result, the inefficient use of some cores near the end of a compute task was a necessary downside to allow for more efficient packing of those compute nodes at the start of each task. Similarly, while there is the potential to pack more threads per compute node where the memory use does not achieve the peak on multiple threads simultaneously, there was no easy way to predict this---necessitating us to risk being slightly less efficient in the number of sensors handled at once in favor of avoiding losing entire nodes to exceeding the available memory. Finally, we note that in some cases it may have been possible to exceed the number of threads in order to use hyperthreading; this possibility is interesting in practice, but we ultimately found that any perceived gains were minimal---though this is hard to disentangle from the possibility that it is related to load balancing on the CPU.

Our workflow optimizes around these constraints with a priority that most compute tasks will be limited to a single node with the maximum number of available sensors given memory constraints, but maintains the ability to pack together different visits as necessary. Each task schematically is just a list of Python commands that need to be run combined with all the relevant input and output paths. In order to avoid repetition of tasks, we added a simple check that can be run between or even during major compute runs which simply validates the location of expected output files in order to determine if a sensor still needs to be simulated. A Parsl app was written to execute each list of commands inside of a container. Figure~\ref{fig:imsimlayout} provides a schematic look at the overall flow from start to finish.

\section{Results}
\label{section:results}
We divide our discussion of the results into three areas. To begin, we describe production runs of the \name{} focusing on  how much data was produced and how much compute time was consumed. 
We then investigate the overheads and system utilization of our approach. 
We conclude by exploring the scalability of the workflow.

\subsection{Production runs}

The total resources used by the simulation can be broken up into two categories. The first is the WFD survey that was carried out exclusively using Cori at NERSC. This effort scaled up to 2000 total compute nodes at any given time and ultimately consumed $\approx 90$M core-hours to fully process the $\approx 2.6$ million images in the survey volume.

The simulation of the DDF survey was carried out utilizing Theta at ALCF. While previous pathfinding jobs were carried out on upwards of 4000 nodes on Theta, the final production run was limited to a maximum of 2863 nodes---this number was ultimately a factor of the available work in natural breaks of the data combined with perceived better scheduling. Note that the DDF survey simulation possesses a natural computational edge in that it is capable of building off checkpoint files done at other facilities; it functionally only needs to draw additional objects onto the images that were not necessary for the WFD survey. As a result, this additional area only required $\approx 10$M compute hours at Theta in order to complete.

\subsection{Overhead}
While introducing additional overhead via containerization may be a concern, we note that task bundling, worker distribution, loading of the containers, and loading of necessary Python libraries takes no longer than 15 minutes on average on Theta---a fairly low cost in comparison to the code run time of around ten hours. In contrast, it has been noted that loading large Python libraries across many nodes simultaneously can result in file system slowdown in a number of other use cases; it is quite possible that even without a container, we would see comparable load times.

While there is a potential for some performance loss due to the use of containers, 
the convenience of being able to run on multiple computing resources outweighs these concerns. Indeed, anecdotally, similar workflows on these resources have demonstrated that the performance loss is minimal.

\subsection{System utilization}
\begin{figure}[h]
  \includegraphics[width=0.97\columnwidth]{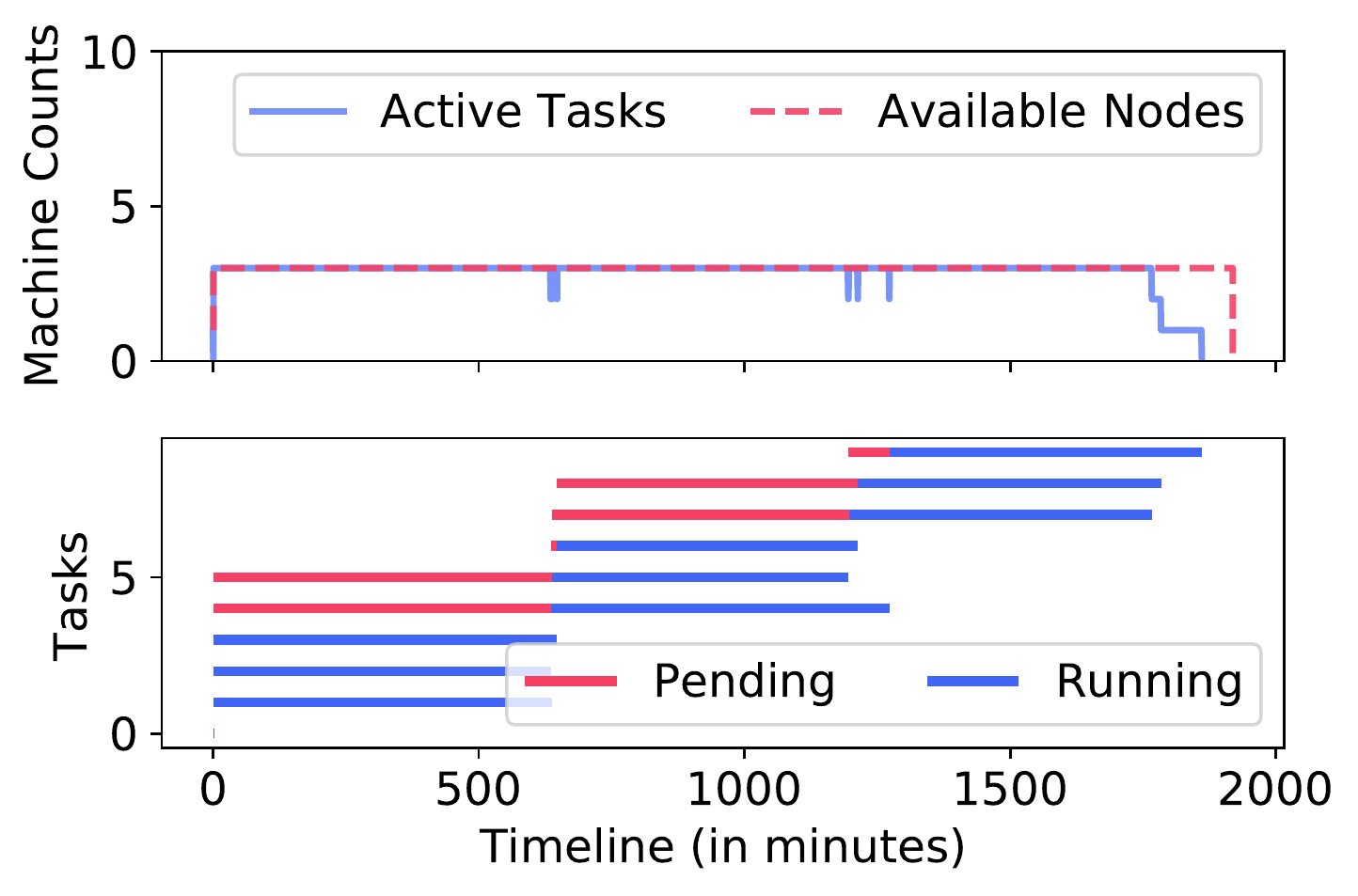}
    \vspace{-0.1in}
  \caption{\name{} machine utilization for a small-scale experiment.  The upper figure shows the total active tasks vs available nodes over time. The lower figure shows individual tasks and their status color-coded, along time.}
  \label{fig:utilization}
    \vspace{-0.1in}
\end{figure}

We conducted a small-scale experiment to observe 
fine grain system utilization. Here, we submitted 
a subset of the full simulation and provisioned three
nodes on the Cori supercomputer KNL nodes. We see in Figure~\ref{fig:utilization} that the submitted task bundles
run for approximately 10 hours each, and consume
most of the available resources.  We also see that
tasks remain queued until there is available 
capacity, and that some tasks are only created
when other tasks have completed. Some underutilization of compute nodes can be noted when the number of remaining tasks goes to zero---this is driven by the significant differences in compute time from image to image and no longer being able to fill the nodes with new work. While this experiment was carried out on a small number of nodes, we observe similar behavior at several thousand compute nodes.

\subsection{Scalability}
With regards to scaling of {\tt imSim}, the division of tasks at the level of sensors proves to be bounded primarily by the amount of available memory on a compute node. With 98 GB of RAM available per node on Cori, this ends up being a safe upper limit of 33 sensors per compute node. The 196 GB of RAM available per node on Theta results in this number being raised to 64, primarily due to evidence that hyperthreading does not seem effective at reducing compute time. It should be noted that this scaling is not perfect --- while direct comparisons are difficult, the compute needs of the image simulation are large enough that increasing the number of active threads on a processor can result in reduced clock speed due to overall power consumption. On Theta, we achieved peak use of the machine at running with 4000 compute nodes simultaneously. This results in a peak usage of 256,000 images being simulated, handled by 4000 Parsl workers running across 4000 compute nodes. While traditional scaling measurements are difficult due to the combination of the computational cost and the high image to image variance, the Parsl executor scaling has been demonstrated in other works \cite{babuji2019}.

%-------------------------
\section{Discussion}
\label{section:discussion}
%-------------------------
The \name{} served to enable a major simulation effort, scaling up to thousands of nodes and allowing us to finish production in our target time frame. The success of this workflow demonstrates the viability of Python-based parallel workflows and use of containers in support of a complex simulation; since this work several new workflows
have been developed following similar methods, but with significantly different codes and goals. With that in mind, we describe ways to improve our
approach for future simulation efforts.

One particularly exciting area to consider is the use of remotely driven workflows. Workflow drivers could even operate on resources entirely external to the HPC facility, such as a user's laptop or in the cloud. In our simulation effort, we favored an approach in which the Parsl workflow driver ran on compute or launch nodes in order to avoid workflow disconnects at critical junctions; this disconnect could be the result of scheduled maintenance, a login node failure, or a network connection failure as some common examples. At the time of our production, this required a manual reconnect of the drivers to worker tasks, which may not be always feasible depending on when jobs are started.

In the case that the workflow driver can be safely hosted on a stable environment, there are a number of improvements to computational and practical efficiency that can be made. One advantage is that some compute facilities potentially favor running jobs in smaller blocks; having the driver running on the compute node can allow simulation tasks to be easily divided even among separate jobs on the compute nodes. This also potentially reduces the amount of wasted compute as the simulation tasks are exhausted---rather than one large job in which many nodes have little work, one will likely be left with only a single small job. As many facilities are starting to explore workflow nodes that are designed for hosting drivers in such a way, this becomes an increasingly promising direction.

A more ambitious possibility is to consider running the workflow driver external to the computing facility itself, either on cloud services or local clusters. This external site can manage a list of all computational tasks, request resources at multiple compute facilities, and coordinate the distribution of this work based on an assessment of remaining compute allocations, expected wait and wall times, and required resources. As our process uses containers extensively that function across sites, such an approach could be valuable as the compute requirements for tasks continue to grow. However, it may ultimately be limited by specific site restrictions; as one example, Theta compute resources do not have inbound network connectivity and thus would require an agent to be pre-deployed to the system (e.g., using funcX~\cite{chard20funcx}).

Another possible point to consider is the containerization process itself. We briefly alluded above to the fact that the Docker image used in the workflow contains all the underlying code necessary for the simulation. While this has distinct benefits from a portability perspective (as the code can now be run on any facility that supports containers), we note that many compute facilities have undergone significant work to compile optimized Python libraries for their specific architecture. While we did not measure the impact of this on the image simulation code, it might be important to consider this situation in future efforts. At some level, there is a natural back-and-forth on the matter of portability---at what point does a potentially small compute speed-up outweigh the ease of portability and the requirement of reproducibility across multiple platforms.

Finally, there is also the possibility of improvements at the level of the underlying simulation code. The version of {\tt imSim} utilized during the DC2 simulation effort was limited to each sensor simulating the image on an object by object basis, with parallelism limited to the level of processing multiple sensors for a given visit. Current development for {\tt imSim} is exploring an alternative approach where rather than parallelizing on a per sensor basis, the code instead parallelizes on a per object basis---an entire compute node would be occupied drawing objects on a single sensor simultaneously. While the method is still in early stages of exploration, estimates of the per object time to compute suggest that it might help reduce the disparity in compute task run times. This may represent a significant utilization improvement, although one that may have to be in concert with other memory use improvements.

\section{Related Work}
\label{section:related}

The {\tt imSim} package builds on the underlying GalSim library. While this effort represents the first time that a survey was simulated at this scale (with prototype efforts being a factor of 50 smaller), it is likely that other future efforts in astronomy will benefit from the improvement of these simulation methods. Both ESA's Euclid mission and NASA's Nancy Grace Roman Space Telescope have been discussed as benefiting from testing analysis based on GalSim products\cite{rowe2015galsim}. While LSST remains the focus of massive surveys at the moment, it is unlikely to be the last for the astronomy community.

There have been various other efforts
to implement large-scale cosmology
workflows. For example, the LSST
project is developing a scalable
workflow infrastructure in Python. 
Our approaches are compatible, and we are working towards a Parsl-based implementation of the LSST processing 
pipelines. Prior work has also focused
on developing tightly-coupled 
frameworks for analysis of cosmological surveys. One example, 
the Hardware/Hybrid Accelerated Cosmology Code (HACC)~\cite{hacc}, 
supports extreme-scale analysis on supercomputers and can efficiently
make use of accelerators.

There are a large number of workflow systems that could be used to implement 
the \name{}, including 
Pegasus~\cite{pegasus}, Galaxy~\cite{galaxy}, Swift~\cite{swift}, NextFlow~\cite{di2017nextflow}, FireWorks~\cite{jain2015fireworks}, Apache Airflow~\cite{airflow}, and Luigi~\cite{luigi}. 
Pegasus and Galaxy implement a static DAG model in which users define a 
DAG, using an XML document or GUI, and subsequently execute that DAG. Similarly, 
systems like Swift and NextFlow implement a custom domain specific language (DSL). 
We choose here
to instead represent our workflow in Python to support ease of development. 
Like Parsl, Python-based workflow systems such as FireWorks, Airflow, and Luigi
enable specification in Python. However, these systems do not provide
the flexibility required by our workflow to support both local threaded 
computations and remote computations. Nor do they provide hooks for the fine-grained
scheduling implemented in the \name{}.

%----------------------
\section{Conclusions}
\label{section:conclusions}
%----------------------
We have described the development and execution of a high performance simulation workflow to create a large-scale data release of synthetic sky images. Specifically, we have described how we scaled the workflow from managing many individual single-node compute jobs to one that can be easily ported around multiple HPC facilities. While the particular computational limitations of our image simulation workflow may not be applicable to every large simulation effort, we believe that this type of computational effort will become increasingly common going forward in scientific computing. We note that we were able to scale up this code to 256,000 simultaneously running processes on Theta, keeping workers active with compute tasks continuously. The combination of Parsl and containers allowed us to develop a straightforward workflow in Python that scales to high node counts while simultaneously being highly portable.

There are several improvements that could be made at the workflow level through integration with new facility tools. Of particular interest is the usage of ``workflow nodes,'' dedicated and highly-available nodes where we could run the control part of our workflow, which would allow us to utilize multiple submission blocks simultaneously with Parsl naturally farming out the necessary compute tasks without worrying about unnecessarily loading log-in nodes or struggling with various maintenance periods that are necessary for keeping HPC resources running. Additionally, while Parsl provides extensive task monitoring, we found points where additional information is necessary to ease debugging of fringe failure cases. These requirements have motivated the development of improved monitoring capabilities
in recent Parsl releases.

As scientific efforts continue to focus on extremely high precision analysis, the desire to test software driven pipelines in advance of hardware acquired data will persist. Workflows like those we have developed will become an increasingly powerful tool for scientific research. The \name{} demonstrates a way to easily scale up a simulation code to extremely large node counts while remaining highly flexible to leveraging multiple computing facilities. With further improvements, it is entirely possible that future simulation efforts could achieve even higher throughput, improving the ability for scientific pipelines to be robustly tested.

\section*{Acknowledgments}
We thank Salman Habib for carefully reading and editing the manuscript. We thank Ben Clifford for his work developing and debugging the \name{}.

ASV was responsible for implementing the workflow on NERSC and ALCF resources and contributing to the text of the manuscript.
YB supported the development of the workflow and contributed to experiments included in the manuscript.
TU supported the deployment of the workflow on ALCF resources. 
DSK is a co-PI of the Parsl project and contributed to the structure and text of the manuscript.
KC is the PI of the Parsl project and contributed to the workflow architecture and text of the manuscript.  
KH was deeply involved in the overall DC2 effort from initial conceptualization onward and contributed to the manuscript. 

This research used resources of the Argonne Leadership
Computing Facility, which is a DOE Office of Science User
Facility supported under Contract DE-AC02-06CH11357. 
LSST DESC uses the resources of the the
National Energy Research Scientific Computing Center, a
DOE Office of Science User Facility supported by the Office
of Science of the U.S. Department of Energy under contract
No. DE-AC02-05CH11231.
Parsl is supported by NSF 1550588. 

The work of ASV and KH at Argonne National Laboratory was supported under
the U.S. DOE contract DE-AC02-06CH11357.

\bibliographystyle{unsrt}
\bibliography{conference_101719}
\vspace{12pt}
\end{document}